\documentclass[lettersize,journal]{IEEEtran}
\usepackage{amsmath,amsfonts,amssymb,gensymb}
\usepackage{graphicx}
\usepackage{algorithmic}
\usepackage{array}
\usepackage[caption=false,font=normalsize,labelfont=sf,textfont=sf]{subfig}
\usepackage{textcomp}
\usepackage{stfloats}
\usepackage{url}
\usepackage{verbatim}
\usepackage{graphicx}
\usepackage[utf8]{inputenc}
\usepackage[T1]{fontenc}
\def\BibTeX{{\rm B\kern-.05em{\sc i\kern-.025em b}\kern-.08em
    T\kern-.1667em\lower.7ex\hbox{E}\kern-.125emX}}
\usepackage{balance}
\begin{document}
\title{Application of FPGA based Lock-in amplifier for ultrasound propagation measurements using the pulse-echo technique}
\author{Stanisalw Galeski, Rafał Wawrzy\'{n}czak, Claudius Riek and Johannes Gooth
\thanks{S. Galeski, R. Wawrzy\'{n}czak and J. Gooth are with Max Planck Institute for Chemical Physics of Solids, Nöthnitzer Straße 40, 01187 Dresden, Germany.}
\thanks{S. Galeski and J. Gooth are with Physikalisches Institut, Rheinische Friedrich-Wilhelms-Universität Bonn, Nussallee 12, 53115 Bonn, Germany.}
\thanks{C. Riek is with Zurich Instruments Germany GmbH, Mühldorfstrasse 15, 81671 München, Germany.}}


\maketitle

\begin{abstract}
\noindent We describe application of a state-of-the art digital FPGA-based Lock-In amplifier to measurements of ultrasound propagation and attenuation at fixed frequency in low temperatures and in static magnetic fields. Our implementation significantly simplifies electronics required for high resolution measurements and allowing to record the full echo train in single measurement and extract changes in both phase and amplitude of an arbitrary number of echa as a function of an external control parameter. The system is simple in operation requiring very little prior knowledge of electrical engineering and can bring the technique to a broad range of solid state physics laboratories. We have tested our setup measuring the magneto-acoustic quantum oscillations in the Weyl semimetal NbP. The results are directly compared with results previously obtained using standard instrumentation.
\end{abstract}


\section{Introduction}

\noindent Measurement of the speed of sound is one of the most powerful techniques available to solid state physicists. Since the speed of sound is directly related to the materials elastic modulus it is a thermodynamic probe and thus is highly sensitive to phase transitions and quantum oscillations~(QO). In addition, measurement of phonon attenuation provides a handle for the study of dissipation mechanisms complementary to usually employed thermal and charge transport experiments~\cite{1,2}. Indeed, to this day ultrasound techniques have been employed in the study of a wide range of materials from low dimensional~\cite{3,4} and frustrated magnetic insulators~\cite{5} to heavy fermions~\cite{6,7}. In addition, ultrasound measurements are one of the few techniques that can be used in the study of Dirac and Weyl semimetals where other thermodynamic probes often lack sensitivity due to the extremely small charge carrier densities present in these systems~\cite{8,9,10,11}.

In a typical pulse-echo experiment two piezoelectric transducers are fixed to opposite parallel faces of the sample. Subsequently a radiofrequency~(RF) pulse is applied to one of the transducers producing a pulse of acoustic waves at MHz frequencies, that travels through the sample towards the second transducer. On arrival to the second end of the sample part of the mechanical energy is transformed back into an electrical signal by the second piezoelectric transducer with the remainder being reflected back into the sample. In a sample with small attenuation the initial pulse can be reflected there and back several times giving rise to a ‘train of echoes’ recorded by the receiving transducer. In a typical measurement the probe pulse is repeated with a kHz frequency allowing to record both the amplitude and time of arrival of the echoes as a function of external parameters such as magnetic field, temperature or mechanical strain~\cite{2}. Although knowledge of the length of the sample allows to determine the absolute speed of sound and attenuation per cm, ultrasound experiments gain a new dimension with the application of phase sensitive techniques. Here, the relative change of amplitude and phase shift of the echoes are recorded allowing to attain relative accuracy of at least one part in 10$^{5}$ in $\frac{\Delta{}v}{v}$~\cite{2}.

Although very powerful, the standard implementation of analogue phase sensitive ultrasound measurements is usually technically complex and requires good knowledge of electrical engineering. In a usual implementation a signal in the $5-500$~MHz range is provided by a frequency generator and then divided into two parts one going to the sample and one serving as a reference signal. The signal is subsequently modulated into pulses, amplified using a power amplifier and send into the input transducer. On arrival to the receiver side, the signal is amplified and multiplied by the reference and a replica shifted by $90$~degrees, and passed through a low pass filter. This is the functionality of a lock-in amplifier.  Subsequently in and out of phase components of the signal are averaged in a narrow time window, usually containing a single echo, using boxcar averagers. Then the output from the boxcar averager is passed to voltage meters and recorded with a computer. The amplitude and phase of the arriving echo are recorded over multiple repetitions of the sequence to improve the signal-to-noise ratio~\cite{12}.
 
This measurement procedure has been significantly simplified by the use of the digital technique. Here the full arriving echo train is directly recorded by a digital oscilloscope and then post-processed using a computer~\cite{13}. Although such approach simplifies the used electronics it puts additional requirements on the data transfer between the oscilloscope and the computer and requires design of specialistic software for data processing. 

Here we propose a further simplification of the measurement utilizing the Zurich Instruments UHFLI lock-in amplifier with an integrated Arbitrary Waveform Generator~(AWG) option and based on FPGA~(Field-Programmable Gate Array) technology. In our implementation the generation of signals, analysis and processing of the input signal are performed with the same instrument. For test measurements we have implemented our measurement setup using the Quantum Design Physical Property Measurement System~(PPMS) DynaCool cryostat. A particular strength of our implementation lies in its simplicity and the possibility to fully automate the measurement via interfacing both the UHFLI and the PPMS DynaCool system within the LabVIEW environment. Our test measurements of magneto-acoustic quantum oscillations in a prototypical Weyl semi-metal NbP yield excellent agreement with previous studies~\cite{9} demonstrating that our approach despite its simplicity achieves the same resolution as more traditional setups.

\section{The instrument}

\begin{figure}[!t]
\subfloat{\label{fig1a}}
\subfloat{\label{fig1b}}
\subfloat{\label{fig1c}}
\subfloat{\label{fig1d}}
\subfloat{\label{fig1e}}
\centering
\includegraphics[width=\columnwidth]{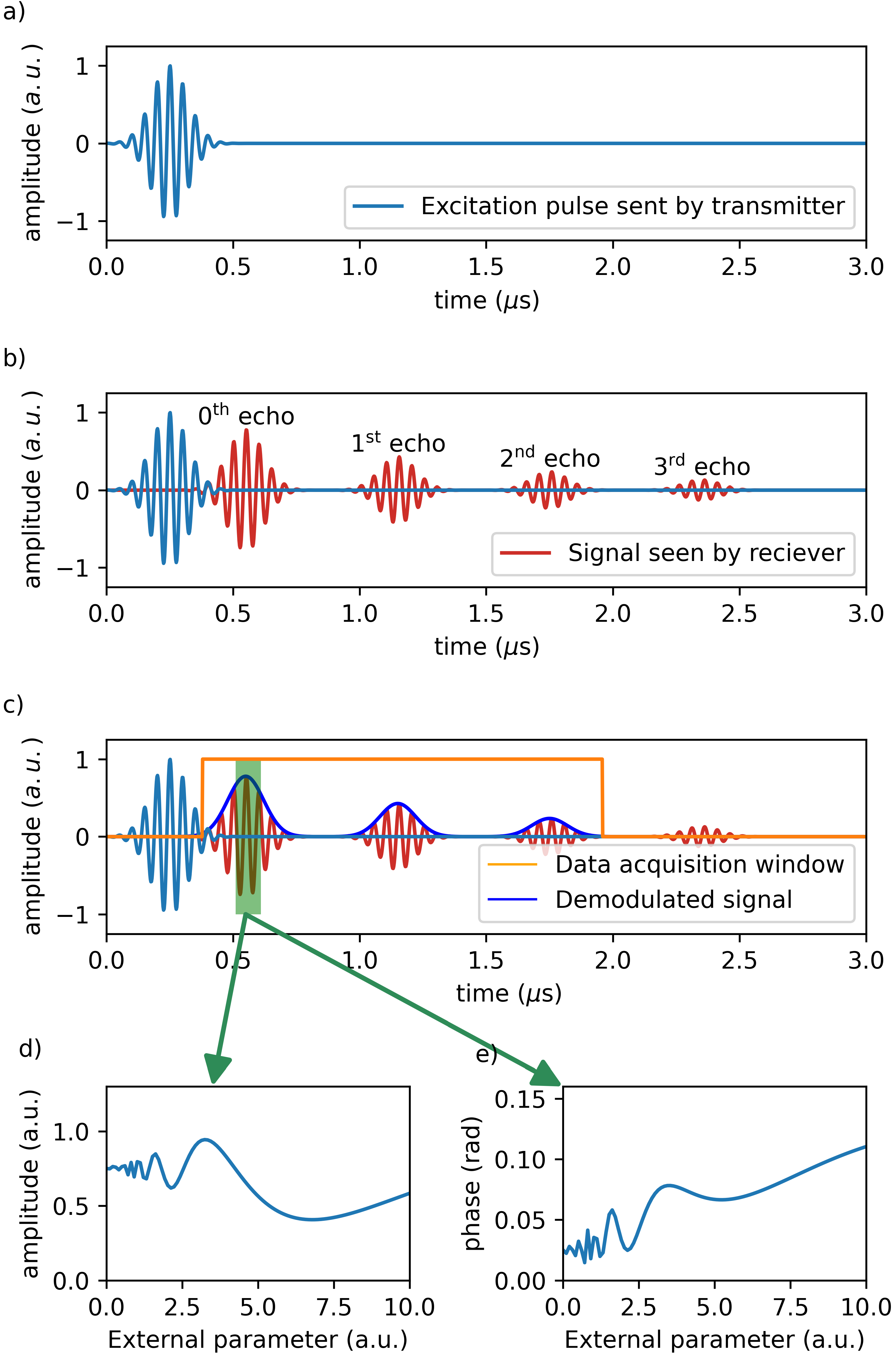}
\caption{Stages of the measurement and data acquisition. a) Excitation pulse generated by AWG. b) Echo train recorded by the receiving transducer. c) Data acquisition window timed with the instrument’s internal clock, and the demodulated signal within this window. d) and e) Change of phase and the amplitude of the 0$^\textrm{th}$ echo, integrated in the range marked by green rectangle, plotted as a function of external parameter.}
\label{fig1}
\end{figure}

\noindent There are three crucial tasks for the ultrasound measurements as described above. (i) Generation of a short excitation pulses at a specific RF-frequency, (ii) demodulation of the response signal from the transducer setup and (iii) separation of the echa in the time domain from the instantaneous EM-crosstalk between the feed lines of the transducers. This results in amplitude and phase measurements per echo relative to the excitation pulses, which is then averaged over multiple pulses sent through the sample. In our implementation all three tasks are performed by the Zurich Instruments UHFLI lock-in amplifier which has the optional AWG built in. The latter generates the excitation pulses at the required RF-frequency (Fig.~\ref{fig1a}) by multiplying the point by point specified envelope with the carrier from a numerically oscillator based on FPGA. Also, the AWG generates the trigger signal~(Fig.~\ref{fig1c}) to isolate the region of interest after the demodulation and controls the synchronized data acquisition from the lock-in amplifier. The delay and length of trigger signal allow to adjust the data acquisition window to enclose desired number of echoes (Fig.~\ref{fig1c}). The lock-in amplifier block of the UHFLI demodulates the input signal arriving from the second transducer with the internal reference signal from the numerical oscillator. Governed by the trigger signal from the AWG, which could also be provided externally. Since both functional blocks of the instrument, the AWG and the demodulators are implemented on the same FPGA they work synchronized down to an individual sample of the processed data stream ($1.8$~GSa/s).  Accordingly, the reference signal for the demodulator is a digital replica of the carrier of the excitation signal sent to the first transducer in the first place. Hence, all disturbances on the signal can only come from the rest of the setup.

\section{Experimental}

\begin{figure}[!t]
\subfloat{\label{fig2a}}
\subfloat{\label{fig2b}}
\subfloat{\label{fig2c}}
\centering
\includegraphics[width=\columnwidth]{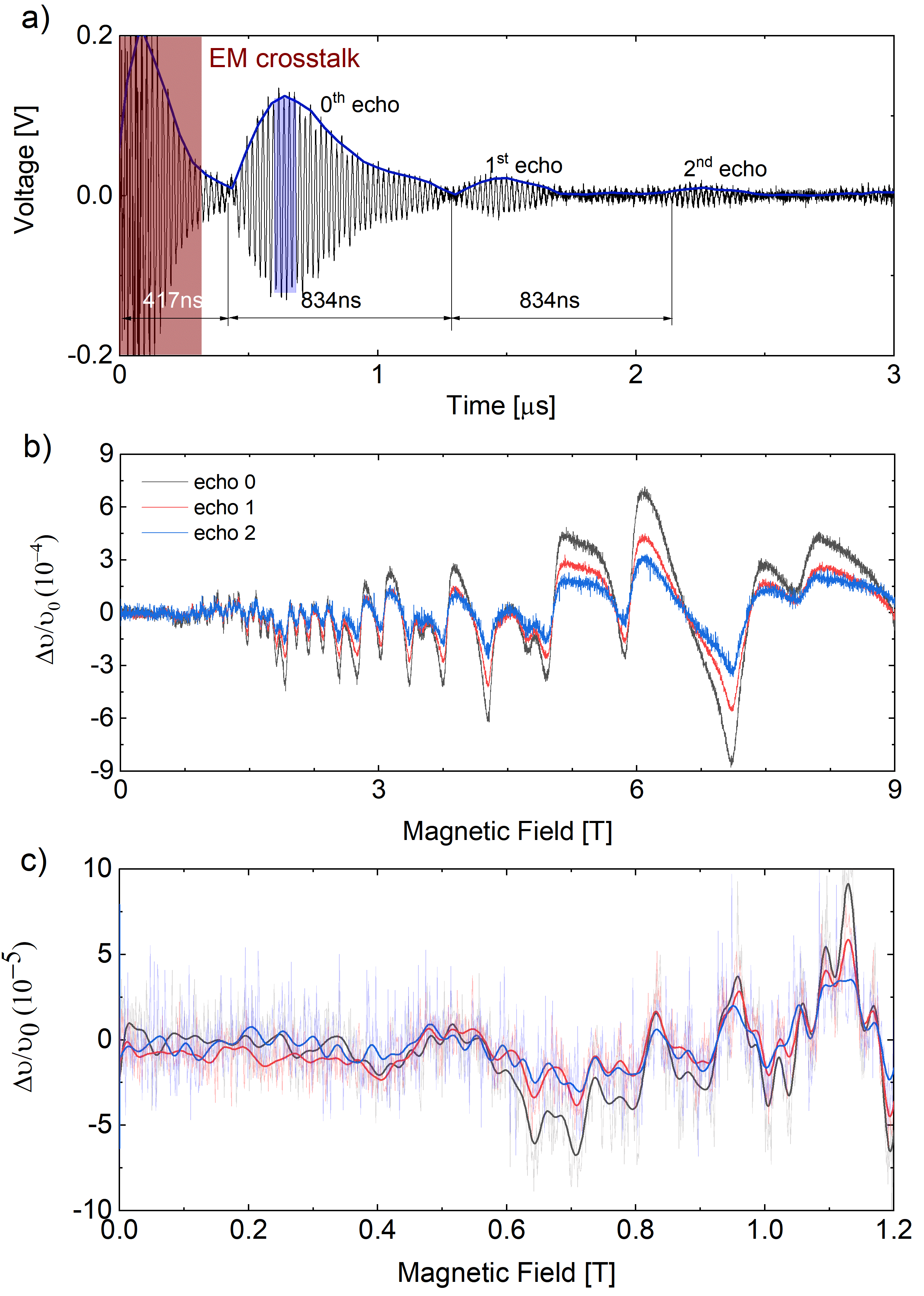}
\caption{a) Raw echo train data recorded using the ZI UHFLI, the solid line represents the demodulated signal. The red shaded region marks the part of the signal contributed to electromagnetic cross-talk The blue shaded region reflecte the part of the demodulated signal used to extract the field dependance of the speed of sound and its attenuation. b) Relative speed of sound of the C33 elastic mode measured using data from different echo measured using 45MHz excitation. c) Zoom in of the low field data. The thin faded lines represent raw measurement data. The thick solid lines represent a 10-point average of the raw signal added for clarity of presentation.}
\label{fig2}
\end{figure}

\noindent In order to test our setup, we have measured propagation of the longitudinal $C_{33}$ sound mode in the semi-metal NbP at low temperatures and in static magnetic fields up to $9$~T. Previous studies have identified NbP as a material realizing Weyl semi-metal~\cite{15}. It is characterized by complex and highly anisotropic Fermi surface, consisting of several hole and electron pockets. The small cross-section of the Fermi surface components allows the observation of pronounced quantum oscillations at relatively small magnetic fields. In particular our recent study revealed a presence of strong magneto-acoustic quantum oscillations already in fields below $10$~T~\cite{9}, making NbP a perfect testing ground for our ultrasonic measurement setup.
The crystal used in this work was grown using the chemical vapor transport methods. NbP grows in a tetragonal crystal structure with the unit lattice parameters $a=b=0.33324$~nm and $c=1.113705$~nm. For the purpose of this experiment the as-grown crystal has been cut using a wire saw with two opposite faces perpendicular to the $[0,0,1]$ crystallographic direction, placed $2.74$~mm apart. The surfaces were fine polished for attachment of the ultrasonic transducers. Afterwards two lithium niobate (LiNbO$_{3}$, $36$\degree Y-cut) transducers were glued to the polished surfaces for excitation and detection of the acoustic waves. In this configuration the probed acoustic waves correspond primarily to the $C_{33}$ longitudinal mode. The measurements were conducted in the PPMS DynaCool system equipped with a $9$~T vertical magnet. The field was applied along the $[1,0,0]$ crystallographic direction. Since the PPMS DynaCool system is not equipped with wiring necessary for the use of radio frequencies a custom insert stick adapted from a commercially available thermal expansion/magnetostriction setup,  equipped with coaxial RF cables, was employed~\cite{14}.

In the test measurements we have utilized a $45$~MHz excitation signal sending $140$~ns long pulses with $1.3$~kHz repetition rate into the sample. In order to amplify the excitation, signal we have used the Amplifier Research 50U1000 power amplifier. Figure~\ref{fig1a} shows an example echo train recorded at $B=0$~T and $T=1.8$~K. Measurement of the time of arrival of the echoes together with the $2.74$~mm sample length allows to estimate the speed of sound to be ca. $v=6.56$~km/s. Using the measured speed of sound and density of NbP ($\rho=6.52$~g/cm$^{3}$) we have estimated the value of the elastic modulus to be $C_{33}=280$~GPa, in good agreement with literature values~\cite{9}. 

\begin{figure}[!t]
\subfloat{\label{fig3a}}
\subfloat{\label{fig3b}}
\centering
\includegraphics[width=\columnwidth]{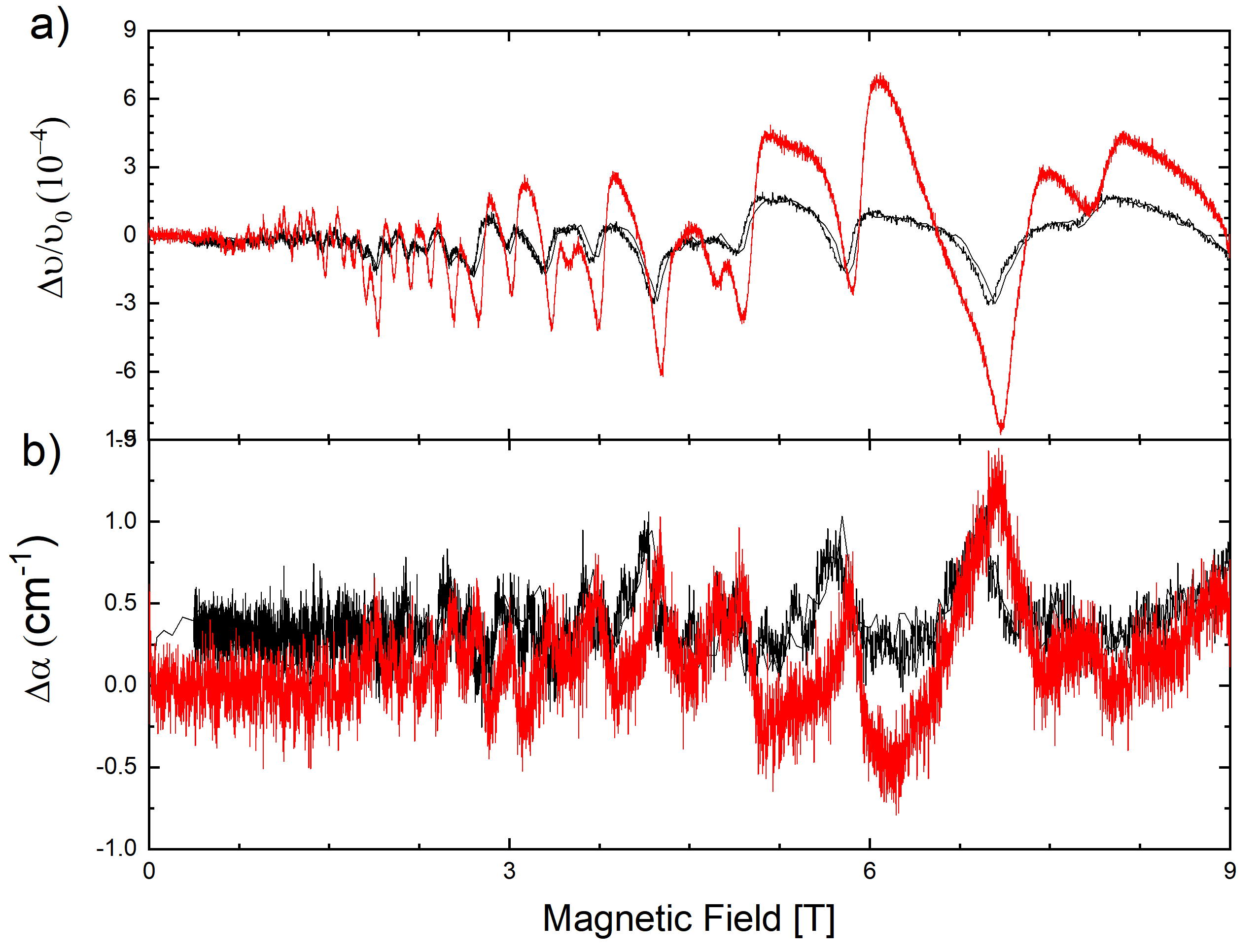}
\caption{Comparison of quantum oscillations of a) sound veolocity and b) ultrasound attenuation in NbP measured using a traditional setup at the HZDR pulsed field facility during a $150$~ms pulse (black) and those obtained using our FPGA based setup measured in a QD Dynacool cryostat (red). Pulsed Field data has been previously published in Ref.~\cite{9}.}
\label{fig3}
\end{figure}

During the measurements in magnetic fields the received signal was demodulated using the Lock-in internal electronics and its phase and amplitude at the maxima of each echo were recorded simultaneously as a function of magnetic field and temperature. Results of test measurement of the relative change of the speed of sound using signal from all three recorded echoes as a function of magnetic field are shown in figure~\ref{fig2b}. In all measurements the field was ramped with a rate of $0.002$~T/s. Data from all echoes consistently shows magneto-acoustic QO in the speed of sound, albeit with data quality slightly deteriorating for higher echo numbers. At first glance due to the large magnitude of the QO at high fields it is difficult to benchmark the resolution of our measurement setup. However, this can be investigated by inspecting the low field region showed in figure~\ref{fig1c}. Here it is clearly seen that all $3$ measurements display reproducible features of minimum size of less than $1$ part in $10^{5}$. For a direct comparison with standard approaches to ultrasound measurements. Figure~\ref{fig3}  displays a direct comparison between data of sound velocity and attenuation measured on our setup and previously collected in a pulsed field experiment in a high field lab of Helmholtz-Zentrum Dresden-Rossendorf~(HZDR), measured on NbP crystals from the same batch. Again, the excellent agreement with previous measurements confirms the reliability of the proposed implementation.  

A particular strength of our implementation lies in its simplicity. Thanks to avoiding excessive cabling and interfacing of multiple instruments, each of which temperature can drift leading to drift in the signal, our setup proved extremely reliable allowing to collect data during several days without experiencing any jumps and glitches. Figure~\ref{fig3a} shows and exemplary raw dataset collected over a period of $72$ hours. As can be seen the measured data is of very high quality without any jumps and shifts. To further test the accuracy of our measurements we have performed a Fourier transform of our data in order to extract the frequencies contributing to the QO. Each frequency representing an extremal electron orbit around of the pockets constituting the Fermi surface. The Fourier spectrum showed in figure~\ref{fig2b} yields excellent agreement with previous studies with the dominant frequencies being $F_{1}=1.6$~T, $F_{2}=7.62$~T, $F_{3}=13$~T, $F_{4}=31.7$~T, $F_{5}=62.9$~T and $F_{6}=94$~T. In addition our analysis of the temperature dependence of the Fourier spectrum allowed to extract the effective cyclotron masses related to each of the orbits with: $m_{c}^{F1}=0.065m_{e}$, $m_{c}^{F2}=0.069m_{e}$, $m_{c}^{F3}=0.085m_{e}$, $m_{c}^{F4}=0.102m_{e}$, $m_{c}^{F5}=0.149m_{e}$, $m_{c}^{F6}=0.206m_{e}$, again in good agreement with previous reports~\cite{9,15}. 

\begin{figure}[!t]
\subfloat{\label{fig4a}}
\subfloat{\label{fig4b}}
\centering
\includegraphics[width=\columnwidth]{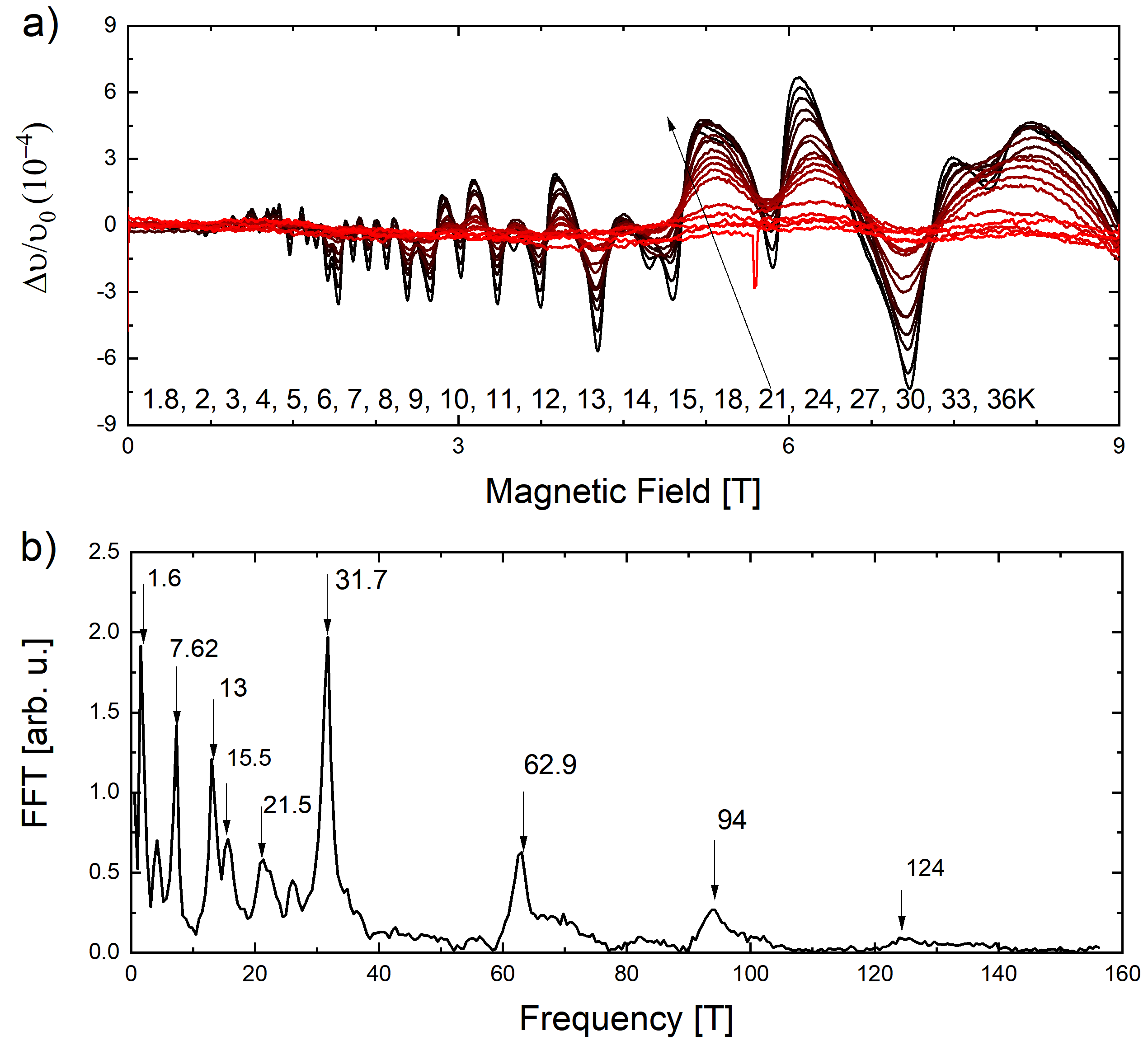}
\caption{a) Temperature dependence of the magneto-acoustic quantum oscillations in NbP b) A Fourier transform of the quantum oscillations. Arrows mark maxima of the spectrum representing particular electron orbits}
\label{fig4}
\end{figure}

\section{Conclusions}

\noindent We have presented the successful application of cutting-edge FPGA-based lock-in amplifier in phase sensitive ultrasound propagation measurements. With the employed method we were able to significantly reduce the number of required electronic components of the measurement setup to the absolute minimum of two (\textit{i.e.} lock-in amplifier and power amplifier) and by that replace the usually used setup filling the whole equipment rack. The data acquisition was performed by interface created in LabVIEW environment, allowing for easily attained compatibility with wide range of sample environment systems (\textit{e.g.} Quantum Design PPMS). The test measurement performed on Weyl semi-metal NbP has shown the ability of our method to resolve relative changes in sound velocity smaller than $1$ part in $10^{5}$. We hope that, the example of presented setup will contribute to increased accessibility of this powerful technique and result in its introduction to the toolbox of the vast number of solid-state research laboratories.

\section{Acknowledgments}
\noindent We would like to thank S. Zherlitsyn for introducing us to the pulse echo measurement technique and for helping to improve the readability of the manuscript. In addition, we also thank M. Schmidt for growing the NbP crystal that we have used for testing of the setup.

\section{Data availability:}
\noindent The data that support the findings of this study are available from the corresponding author upon reasonable request.

\end{document}